\documentclass[a4paper,conference]{IEEEtran}

\ifCLASSINFOpdf
\else
\fi
\ifCLASSOPTIONcompsoc
 \usepackage[caption=false,font=normalsize,labelfont=sf,textfont=sf]{subfig}
\else
 \usepackage[caption=false,font=footnotesize]{subfig}
\fi
\usepackage{url}

\usepackage[english]{babel}

\usepackage{xcolor}

\usepackage{soul}

\usepackage[textsize=tiny,colorinlistoftodos]{todonotes}

\makeatletter
    \define@key{todonotes}{sm}[]{
        \setkeys{todonotes}{author=\textbf{Sergiy}, color=green!30}}%
    \define@key{todonotes}{qz}[]{
        \setkeys{todonotes}{author=\textbf{Qiuheng}, color=blue!30}}%
    \define@key{todonotes}{if}[]{
        \setkeys{todonotes}{author=\textbf{Ingo}, color=lime!30}}%
    \define@key{todonotes}{jk}[]{
        \setkeys{todonotes}{author=\textbf{Jochen}, color=cyan!30}}%
    \define@key{todonotes}{mgs}[]{
        \setkeys{todonotes}{author=\textbf{Mandy}, color=orange!30}}%
\makeatother

\usepackage[range-phrase={\,--\,}, range-units=single, per-mode=symbol, detect-all]{siunitx}

\makeatletter
  \@addtoreset{paragraph}{section}
\makeatother

\usepackage[acronym]{glossaries}
\newacronym{vr}{VR}{Virtual Reality}
\newacronym{xr}{XR}{eXtended Reality}
\newacronym{ar}{AR}{Augmented Reality}
\newacronym{qoe}{QoE}{Quality of Experience}
\newacronym{qos}{QoS}{quality of service}
\newacronym{uav}{UAV}{unmanned aerial vehicle}
\newacronym{faas}{FaaS}{Function-as-a-Service}
\newacronym{pubsub}{Pub/Sub}{Publish/Subscribe}
\newacronym{h3d}{H3D}{holographic 3D}
\newacronym{ai}{AI}{Artificial Intelligence}
\newacronym{nbmp}{NBMP}{network-based media processing}
\newacronym{kpi}{KPI}{key performance indicator}
\newacronym{hc}{\emph{HoloCom}}{\emph{Holographic Communication}}
\newacronym{ml}{ML}{machine learning}
\newacronym{rem}{REM}{radio environment map}
\newacronym{drl}{DRL}{deep reinforcement learning}
\newacronym{sdr}{SDR}{software-defined radio}
\newacronym[firstplural=radio access technologies]{rat}{RAT}{radio access technology}
\newacronym{ran}{RAN}{radio access network}
\newacronym{tcas}{TCAS}{traffic alert and collision avoidance system}
\newacronym{ga}{GA}{general aviation}
\newacronym{mcs}{MCS}{Modulation and Coding Scheme}
\newacronym{fec}{FEC}{forward error correction}
\newacronym{api}{API}{application programming interface}
\newacronym{lstm}{LSTM}{long short term memory}
\newacronym{cnn}{CNN}{Convolutional Neural Network}
\newacronym{cam}{CAM}{Class Activation Map}
\newacronym{sdn}{SDN}{Software Defined Network}
\newacronym{vm}{VM}{Virtual Machine}
\newacronym{ui}{UI}{User Interface}
\newacronym{db}{DB}{Data Base}
\newacronym{iac}{IaC}{Infrastructure as Code}

\usepackage{csquotes}
\usepackage{balance}
\usepackage{cleveref}
\usepackage[all]{nowidow}
\usepackage{url}
\usepackage{longtable}

\usepackage[
        bibstyle=ieee,
        citestyle=numeric-comp,
        maxcitenames=6,
        doi=false,
        eprint=false
    ]{biblatex}
        
\AtEveryBibitem{
  \clearfield{urlyear}%
  \clearlist{language}%
 }

\graphicspath{{figures/}}
\begin{document}
%


\title{Over-the-Top Resource Broker System for Split Computing
\\ {\Large{An Approach to Distribute Cloud Computing Infrastructure}
}%
}



%
\author{
    \IEEEauthorblockN{
        Ingo Friese\IEEEauthorrefmark{1},
        Jochen Klaffer\IEEEauthorrefmark{1},
        Mandy Galkow-Schneider\IEEEauthorrefmark{1},\\
        Sergiy Melnyk\IEEEauthorrefmark{2},
        Qiuheng Zhou\IEEEauthorrefmark{2},
        Hans Dieter Schotten\IEEEauthorrefmark{2}\IEEEauthorrefmark{3}
    }
    \IEEEauthorblockA{
        \IEEEauthorrefmark{1}
        Deutsche Telekom AG, Berlin, Germany \\
        Email: \{ingo.friese, jochen.klaffer, mandy.galkow-schneider\}@telekom.de
    }    
    \IEEEauthorblockA{
        \IEEEauthorrefmark{2}
        Intelligent Networks,
        German Research Center for Artificial Intelligence (DFKI), Kaiserslautern, Germany \\
        Email: \{sergiy.melnyk, qiuheng.zhou, hans.schotten\}@dfki.de
    }
    \IEEEauthorblockA{
       \IEEEauthorrefmark{3}
       Institute for Wireless Communication and Navigation,
       University of Kaiserslautern (RPTU), Kaiserslautern, Germany\\
       Email: schotten@rptu.de
    }

    
}

\maketitle


\begin{abstract}
6G network architectures will usher in a wave of innovative services and capabilities, introducing concepts like split computing and dynamic processing nodes. This implicates a paradigm where accessing resources seamlessly aligns with diverse processing node characteristics, ensuring a uniform interface. In this landscape, the identity of the operator becomes inconsequential, paving the way for a collaborative ecosystem where multiple providers contribute to a shared pool of resources. At the core of this vision is the guarantee of specific performance parameters, precisely tailored to the location and service requirements.
A consistent layer, as the abstraction of the complexities of different infrastructure providers, is needed to simplify service deployment. One promising approach is the introduction of an over-the-top broker for resource allocation, which streamlines the integration of these services into the network and cloud infrastructure of the future.
This paper explores the role of the broker in two split computing scenarios. By abstracting the complexities of various infrastructures, the broker proves to be a versatile solution applicable not only to cloud environments but also to networks and beyond. Additionally, a detailed discussion of a proof-of-concept implementation provides insights into the broker's actual architectural framework.
\\
\end{abstract}


\section{Introduction}
\label{sec:introduction}

\begin{refsection}
    \let\thefootnote\relax\footnote{This is a preprint of a work accepted but not yet published at the 29. ITG-Fachtagung Mobilkommunikation. Please cite as: \fullcite{friese2025over}.}
\end{refsection}%



The 6th generation of mobile networks is facing a wide range of new requirements and expectations, leading to a complex, primarily software-based, and cloud-native architecture. The research project \emph{6G NeXt}~\cite{6gnext, 6g2} aims to develop a \emph{Broker Service Framework} that provides services dynamically with resources such as cloud and connectivity. Cloud systems, in particular, are the focus of the framework because \emph{6G~NeXt} implements a dynamic distribution of computing tasks to different available cloud resources in real-time, also known as \emph{split computing}~\cite{melnyk20236gnext}. Future mobile network architectures are positioned to support split computing for \gls{vr} and \gls{ar} applications~\cite{holo} as well as for \gls{ai} services~\cite{bakhtiarnia2022dynamic}. Many devices lack the necessary CPU/GPU power and memory to render complex virtual scenes in near real-time. However, computing tasks can be offloaded to remote computing or rendering services running on edge cloud infrastructure located near the client. In release 18 of its 5G specification, \emph{3GPP} standardized an \emph{Architecture for Enabling Edge Applications}~\cite{3gpp.23.558}. The objective of this architecture is to host applications in an edge cloud close to the base station and thus nearer to the clients, reducing end-to-end latency. The focus of this work is a broker system that extends the 3GPP approach by distributing tasks to various cloud resources—including edge, fog, or central—and across diverse connectivity types.

This paper is structured as follows:
Section~\ref{sec:background} provides background information and illustrates two use cases to describe the problem.
Section~\ref{sec:relatedwork} describes existing concepts and explains how the proposed approach relates to them.
Section~\ref{sec:brokersystem} elaborates on the proposed broker system, including its requirements and core functionalities, as well as the insights gained from a proof of concept implementation and future plans for extensions.
Finally, Section~\ref{sec:conclusionText} summarizes the characteristics of the proposed broker system.

\section{Background and Motivation}
\label{sec:background}

To provide a more comprehensive explanation, we will use two example use cases from the \emph{6G NeXt} project: \emph{Volumetric Video Chat} and \emph{Smart Drones}. 

\begin{figure}[b]
    \centering
    \includegraphics[width=\linewidth]{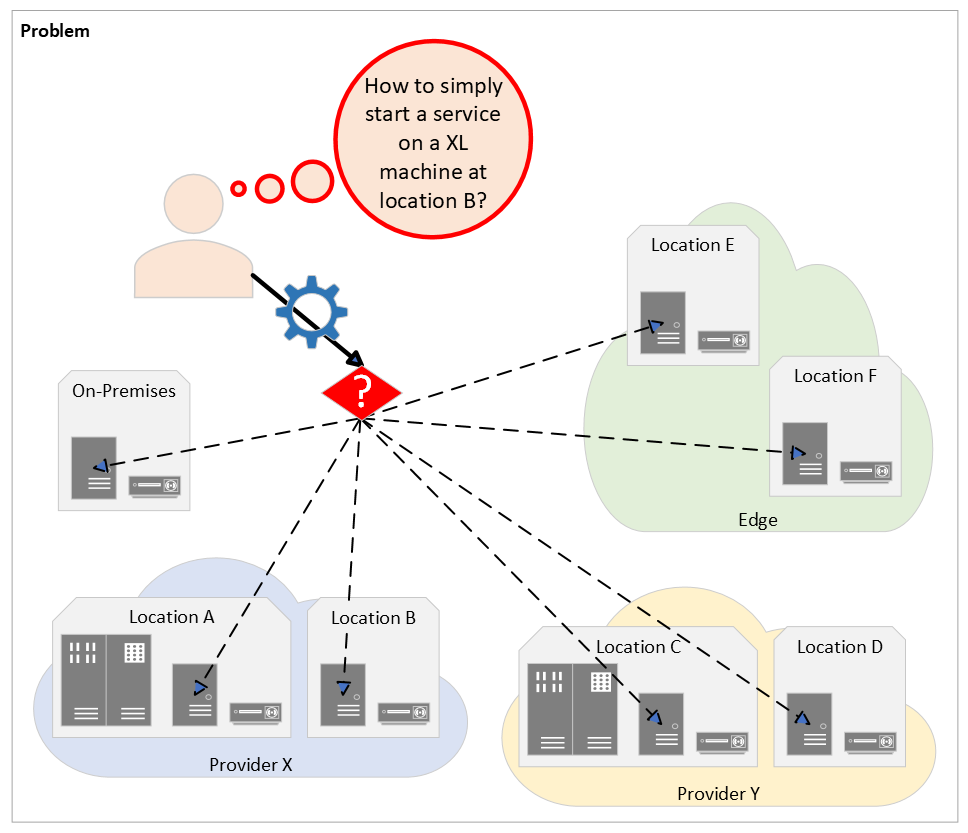}
    \caption{Problem: Services are not meant to handle infrastructure.}
    \label{fig:problem}
\end{figure}

\subsection{Use-Case: Volumetric Video Chat}

In a \emph{Volumetric Video Chat} scenario, local client resources may not be sufficient. This makes it necessary to deploy an additional rendering instance on a nearby high-performance edge cloud. As a result, there is a need to manage the dynamic creation and removal of resources for services or applications, as well as adjust existing infrastructure components. Figure~\ref{fig:problem}~illustrates the problem of a service provider that wants to deploy and start a new service instance.

\subsection{Use-Case: Smart Drones}

\begin{figure}[b]
    \centering
    \includegraphics[width=\linewidth]{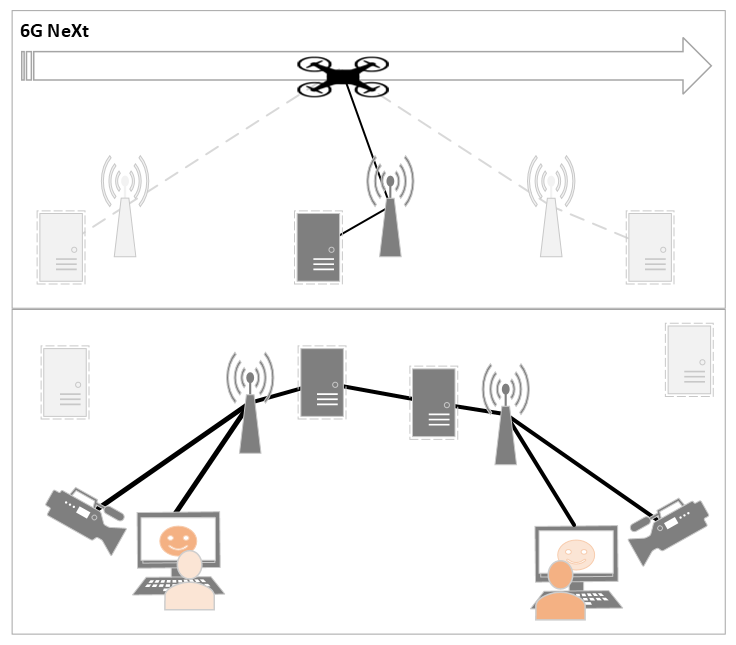}
    \caption{6G NeXt Scenarios: Smart Drones and Volumetric Video Chat running on the edge closest to the client.}
    \label{fig:6gnext}
\end{figure}

The \emph{Smart Drones} use case involves an anti-collision system for drones and general aviation. The goal in this scenario is to provide surveillance services for drones from the ground as needed, leveraging the nearest cloud resource—such as one at an airfield along the flight path—whenever possible. Additionally, the radio link must be configured to minimize latency while maintaining high operational stability as depicted in Figure \ref{fig:6gnext}. Moreover, the required resources, whether on-premises, at the edge, or in the cloud, should ideally be located as close as possible, irrespective of the provider.

\section*{Resource broker system for network and cloud abstraction}

\begin{figure}[t]
    \centering
    \includegraphics[width=\linewidth]{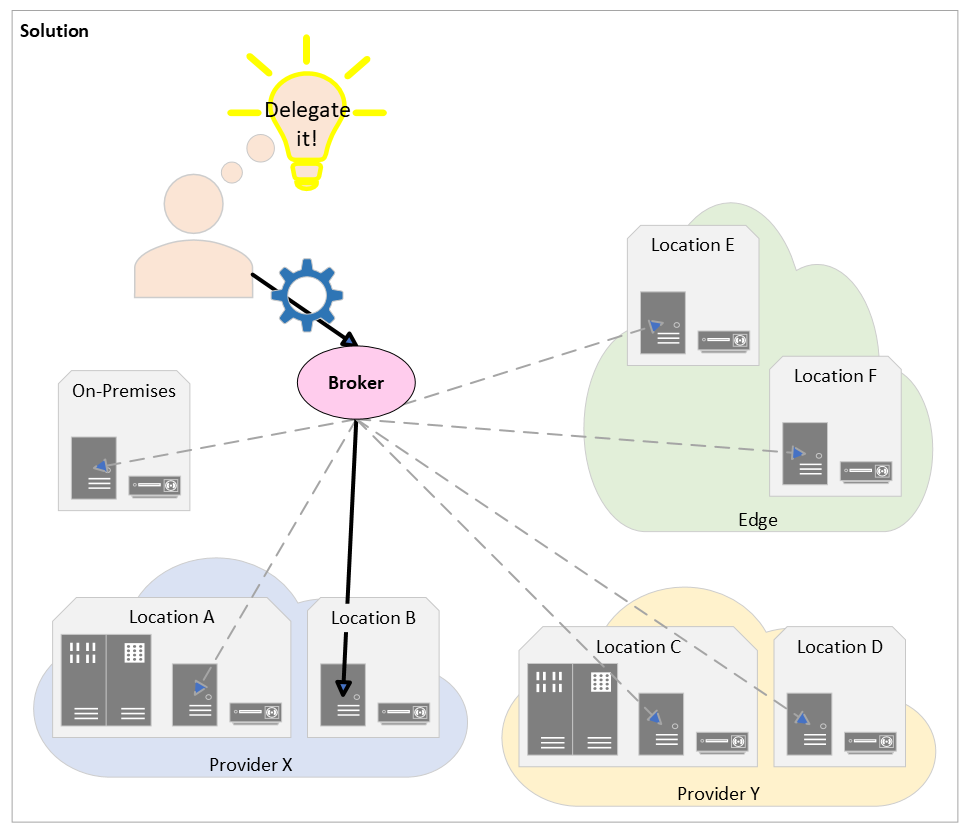}
    \caption{Solution: A resource broker takes over the task of infrastructure preparation for service deployment.}
    \label{fig:solution}
\end{figure}

To relieve services from resource management tasks, we propose an over-the-top resource broker system. This resource broker ensures that the resources required for service operation are efficiently allocated. It establishes a central link between services and the necessary resources, as shown in Figure  \ref{fig:solution}.
This process starts with reserving a virtual machine and extends to managing a connection on a mobile network, covering all intermediate steps. The broker creates a communication channel between the service and resources, delivering tailored solutions and deploying them on specific systems.

The evolution of service architecture—fueled by trends like microservices, containerization, and cloud computing—is increasingly decoupling services from their runtime environments while reducing dependence on specialized hardware. As reliance on specialized hardware decreases, service requirements are shifting toward core computing power, storage capacity, and network throughput.

Service providers today struggle to manage distributed runtime environments across their own data centers and multiple cloud platforms. The diversity of management approaches and technologies adds to this complexity, making provider transitions or replacements during operation highly demanding.

To tackle these challenges, service operators seek a unified, standardized approach to resource provisioning. They require a consolidated resource view that allows them to build tailored runtime environments without managing infrastructure details. While computing performance and network throughput remain key, factors like location, latency, cost, energy efficiency, security, and regulatory compliance are also crucial.

This creates a demand for resource providers that offer a broad range of virtualized resources mapped to real-world equivalents. For example, a virtual machine labeled `XL' in an abstract resource model can be assigned to specific locations and configurations across different cloud providers. This approach simplifies deploying distributed service environments, enabling seamless adaptation and expansion without disrupting operations.

While this provider-centric approach may not offer the flexibility of modern cloud solutions, it simplifies deploying distributed services by offering easily identifiable components. Additionally, the resource broker can provide \glspl{api} for system management, allowing dynamic resource adjustments based on changing needs. Operators can use these \glspl{api} for manual management or as automation interfaces, utilizing \gls{ai} to improve efficiency and adaptability.

\section{Related Work}
\label{sec:relatedwork}

The concept of resource brokers is well-established. Two of the surveys we investigated focus on the contemporary significance of cloud brokering. The first is an article [7], which indicates that cloud brokerage remains a prominent field but still faces technical challenges. Paper [8], on the other hand, explores the economic influences on the use of cloud and edge networks, highlighting the approach of selecting cloud services through a broker, not only based on technical considerations but also on soft factors such as pricing. The authors of [9] introduce a broker system that addresses the growing complexity of cloud infrastructures, enabling dynamic resource provisioning while considering QoS requirements. Additionally, the paper [10] addresses the complexity of cloud service bookings, allowing customers to select providers based on their own criteria.

The broker system discussed in this paper aims to address a broader and more universal scope. New services in areas such as \gls{xr} and \gls{ai} merge computing and connectivity. Therefore, it is important for a broker to facilitate both cloud and network connectivity and select based on desired criteria. These criteria can be expanded beyond cost-effectiveness to include various other categories such as speed, energy consumption, privacy requirements, or jurisdiction (e.\,g. EU or USA), among others.

There are different \gls{iac} tools to manage and provision compute resources through machine-readable definition files, rather than dedicated configurations or interactive configuration tools. \emph{AWS CloudFormation} \cite{AWSCloudFormation} offers a resource modeling service, and tools like \emph{Terraform} \cite{Terraform}, \emph{Google Cloud Deployment Manager} \cite{GoogleCloudDeploymentManager}, \emph{Chef} \cite{Chef}, \emph{Puppet} \cite{Puppet}, and \emph{Ansible} \cite{Ansible} support the management of large environments.

\gls{iac} tools automate the setup, modification, and versioning of infrastructure, processes that can otherwise be time-consuming and prone to errors when done manually. These tools ensure consistency across different environments, reducing bugs and issues caused by configuration drift.
\gls{iac} tools simplify scaling infrastructure to meet growth or changing requirements. Since the infrastructure is defined as code, it also acts as documentation for the setup. In case of an outage or disaster, \gls{iac} makes it easy to rebuild the infrastructure as it defines the entire setup. Changes and updates can be viewed and tracked by teams, making collaboration easier.

However, a significant gap remains in the lack of a methodology for precisely selecting configurations based on descriptive attributes and requirements, as well as the ability to initiate them when needed. This issue will be explored further in future research.

\section{The Broker System}
\label{sec:brokersystem}

\subsection{Requirements and Selection Criteria}

When procuring a \gls{vm} from major cloud providers, decisions typically focus on key technical specifications. Operators must evaluate computing cores, GPU availability, RAM capacity, storage needs, and network connectivity. Cloud providers offer predefined \gls{vm} packages, each with specific configurations, deployment regions and pricing. Despite the similarities among offerings from different providers, direct one-to-one comparisons are often challenging due to variations in package configurations and pricing policies.

\begin{figure}[b]
    \centering
    \includegraphics[width=\linewidth]{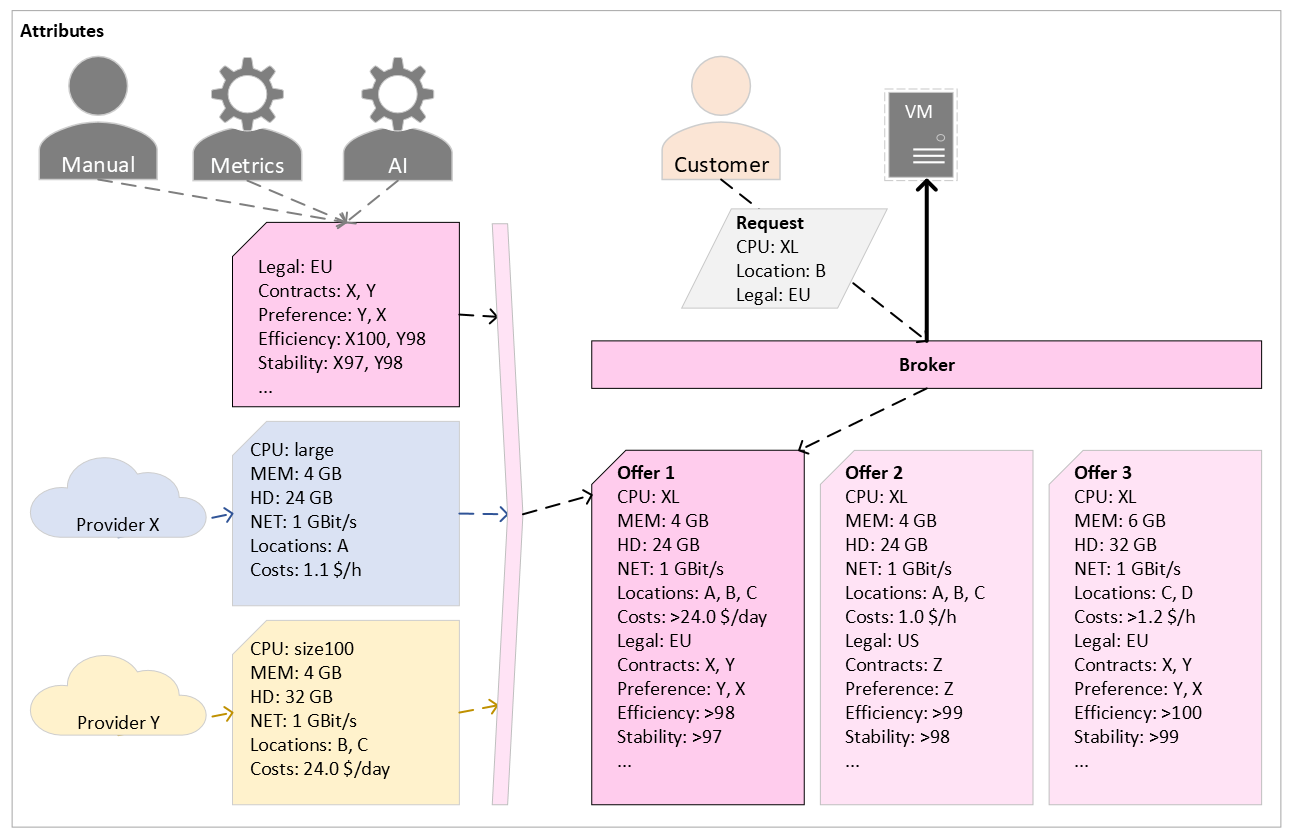}
    \caption{Attributes are the basic building blocks for templates and offers.}
    \label{fig:attributes}
\end{figure}

Beyond core service parameters, selecting a \gls{vm} should also consider ‘soft’ attributes. This requires providers to expand their offerings and provide detailed descriptions, as shown in \Cref{fig:attributes}.

For example, a category like ``efficiency'' whether general or energy efficiency could be introduced, regardless of whether values are estimated, measured, or \gls{vm}-derived. Prioritization could be set, or pricing determined through framework contracts. Legal conditions, such as compliance with EU or US regulations, could also be specified. This serves as a gateway to a broader range of resources, accessible manually, via management applications (policy-driven or AI-supported), or directly through the services.

\subsection{Broker Core}

\begin{figure}[b]
    \centering
    \includegraphics[width=\linewidth]{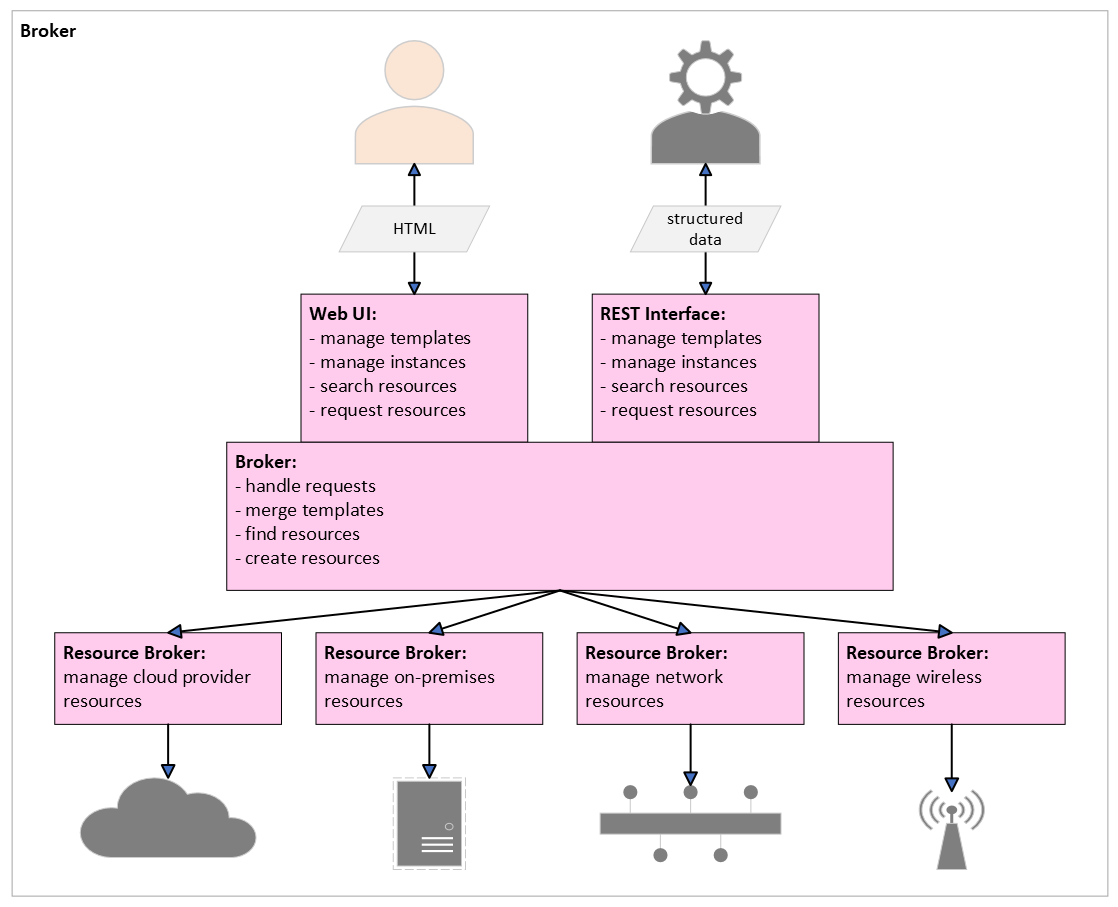}
    \caption{The modules and interfaces of the Broker system.}
    \label{fig:broker}
\end{figure}

The broker is structured as an ordering system, offering the service/customer a REST interface for transmitting requests and a webhook callback mechanism for asynchronous message transmission from the broker to the service. Both messages and replies comprise a straightforward data structure containing embedded user data and a type specification of this user data, ensuring that the communication path remains unaffected when new data types are introduced. Within the messages, several commands are available to interact with the broker, such as ``register VirtualMachine'' or ``remove VirtualMachine'', as depicted in \Cref{fig:broker}. 

To define service offers, the broker compiles a chain of templates, starting with standardized customer templates and concluding with the specific data needed by the underlying providers. The management of these templates is handled through the same REST interface using commands such as register template or remove template, as illustrated in Figure 6. The broker operates based on specific messages, such as:
\begin{itemize}
    \item Command: \texttt{"register"}
    \item Target: \texttt{"VirtualMachine"}
    \item Payload: \texttt{"Data of this VM instance"}
\end{itemize}

Internally, the broker consists of multiple modules or microservices. These components communicate via an event bus and share a common database for storing templates and instances. The broker itself implements processes for communication, resource request implementation, resource instantiation, and management of templates and instances.

\begin{figure}[b]
    \centering
    \includegraphics[width=\linewidth]{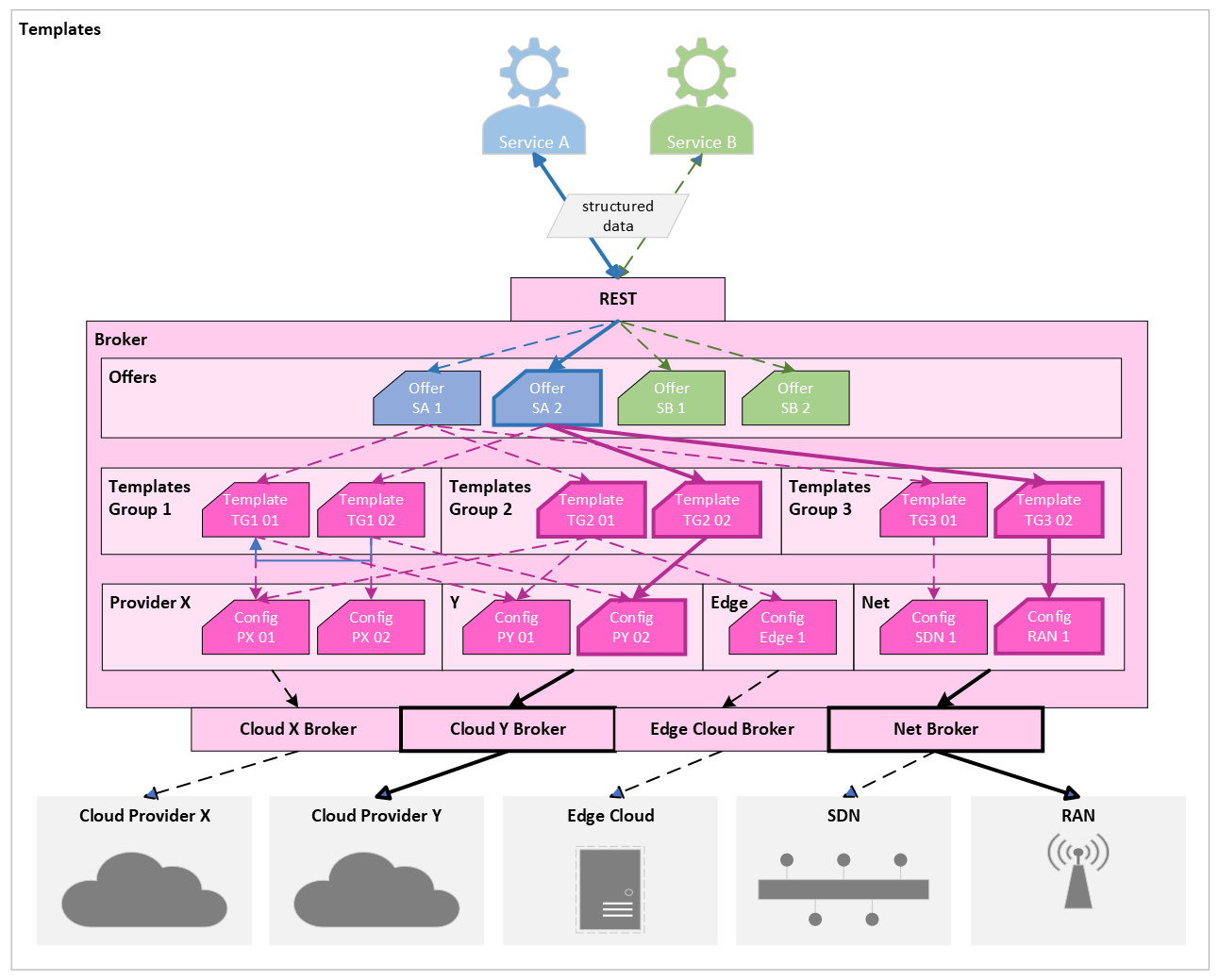}
    \caption{Abstraction through template chains.}
    \label{fig:templates}
\end{figure}

To connect different systems, such as various cloud providers or network segments, specialized modules are required and triggered by specific events. When an incoming ``create'' event is received, it is transformed into an event ``\text{create\_Y}''  to initiate the setup of a resource with the cloud provider ``Y''. These events originate from the templates and are not hard coded within the broker. Therefore, adding a new resource provider is independent, as it requires only a specialized module and corresponding templates, enabling seamless integration into a service offering within the broker.

\subsection{Proof of Concept}

In our Proof of Concept implementation, our focus has been primarily on the initiation and termination of \gls{vm}s across diverse cloud providers and edge cloud deployments.

To achieve this, the broker has been equipped with necessary modules, and offers/templates for the two distinct services have been established. These offers are then aligned with the respective resources of the cloud providers, utilizing various locations offered by the providers. Furthermore, the offers have been enhanced with criteria such as pricing or efficiency.

\begin{figure}[b]
    \centering
    \includegraphics[width=\linewidth]{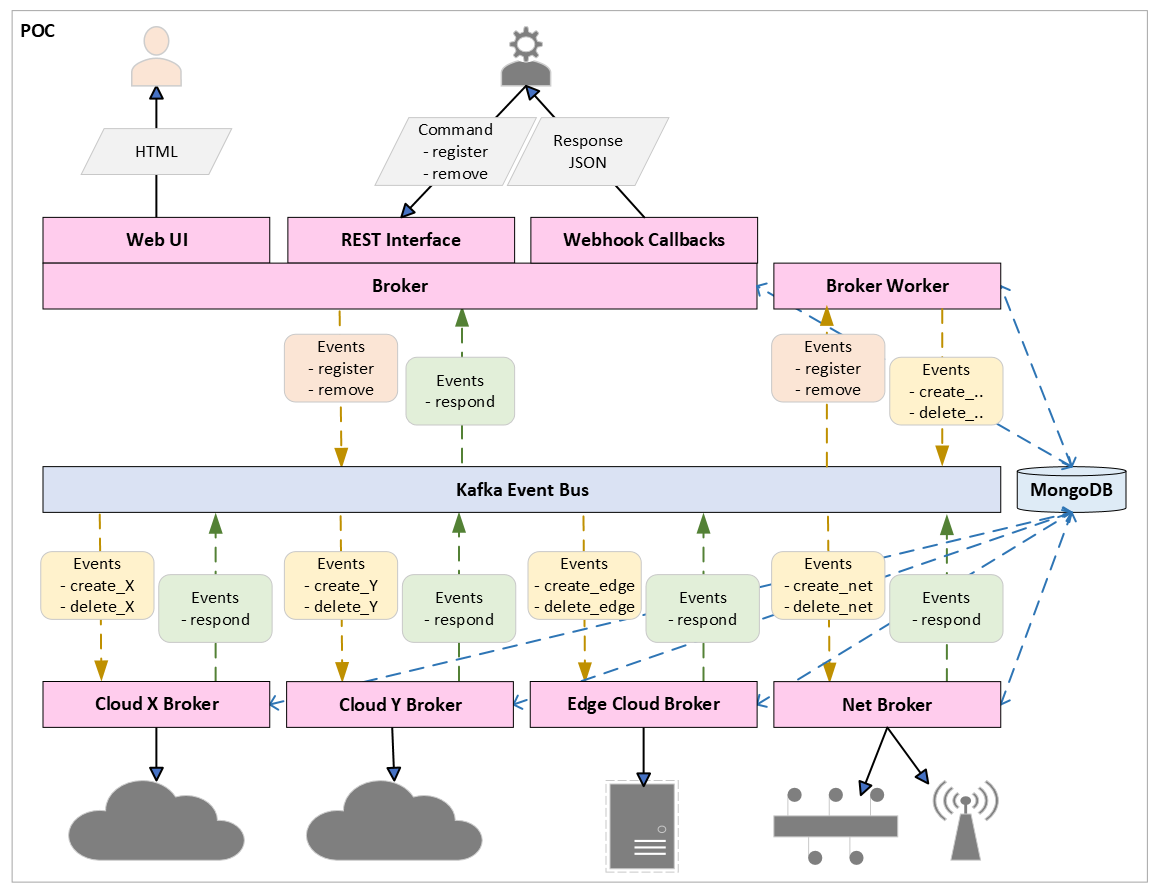}
    \caption{Proof of Concept Setup.}
    \label{fig:poc}
\end{figure}

The current implementation utilizes an event-based architecture, which is realized through microservices and facilitates a data-driven process flow. This architecture consists of a loosely coupled set of functional modules that communicate asynchronously through messages. The sequence of operations is determined by commands and types of messages as visualized in \Cref{fig:poc}. 

All possible commands are defined in the code. To process the data types, suitable modules must be available; otherwise, the messages are discarded. However, new types can be quickly integrated via specific events and corresponding handlers.

Technically, the following specifications have been made: 
\begin{itemize}
    \item The modules are packaged in Docker containers for the Kubernetes runtime environment. 
    \item Kafka is used as event bus and MongoDB as database. 
    \item The modules are encoded in Java using the Spring Boot framework.
    \item JSON is used to structure data.
    \item A basic library is implemented for communication via REST, Kafka, and MongoDB, along with basic event handling and assignment. By a configuration, diverse microservices can be configured to handle various tasks.   
    \item Specialized libraries are accessible for connecting with different cloud and edge cloud providers.
\end{itemize}

Significant effort has been made to keep individual modules stateless, improving the scalability of microservices. Template and instance data are centrally stored in a shared database, ensuring seamless access. For demonstration purposes, a broker service has been developed using these libraries to serve as an entry point for users and services. Additionally, a broker worker has been implemented to manage and distribute incoming requests. Several specialized resource brokers have also been created to handle the connection with individual resource providers. Feedback is sent to requesters via webhooks. Brokers and resource brokers are equipped with a simple web interface to create and manage templates, create or manage resource instances or simply display the required JSON structures for the REST interface in a dry run. In addition, the broker offers a resource search page to list offers filtered by size, location, price and efficiency. 

A possible example of a broker request could be: ``Display all offers of performance class `XL' near the location of the City of Berlin.'' In addition to displaying the resources, it is possible to select and instantiate them directly.

The ultimate goal of the exemplary implementation of the broker is to realize Split Computing scenarios. Within usage scenarios described in \Cref{sec:background}, a cloud rendering service is able to request an additional cloud resource through the broker to start an additional service instance in case of resource shortage. This can be achieved by simply deploying a Docker container and initiating its startup.

In the alternate scenario, a drone control service can request network connectivity with very high reliability, which could be configured using slicing capabilities or network \glspl{api}, such as those specified in CAMARA~\cite{camara}.

\subsection{Outlook}

Up to now, the broker primarily manages \glspl{vm}; we are going to broaden its scope to include lightweight solutions. While Docker containers can currently run alongside \glspl{vm}, integrating an existing Docker runtime remains challenging. Likewise, deploying a microservice into an existing Kubernetes cluster is still unrealized, despite being highly desirable.

Another pivotal aspect of the broker is the management of network resources. In the \emph{6G NeXt} project, our objective is to allocate various channel qualities and bit rates on the radio link. This strategic allocation will empower services to optimize their quality of service through the broker's \gls{api}.

The broker provides an abstraction for uniformly managing underlying resources. The next step is to enable \gls{ai}-driven decision-making and execution, allowing efficient management of network and cloud resources even in complex conditions.

The current implementation manages basic authentication and authorization mechanisms, such as token handling for the underlying infrastructures. However, to fully operationalize the broker approach, these processes need to be enhanced, along with contract and policy management. Ultimately, the broker could form the foundation for a business as an over-the-top provider of cloud and network services

\section{Conclusion}
\label{sec:conclusionText}

\begin{figure}[b]
    \centering
    \includegraphics[width=\linewidth]{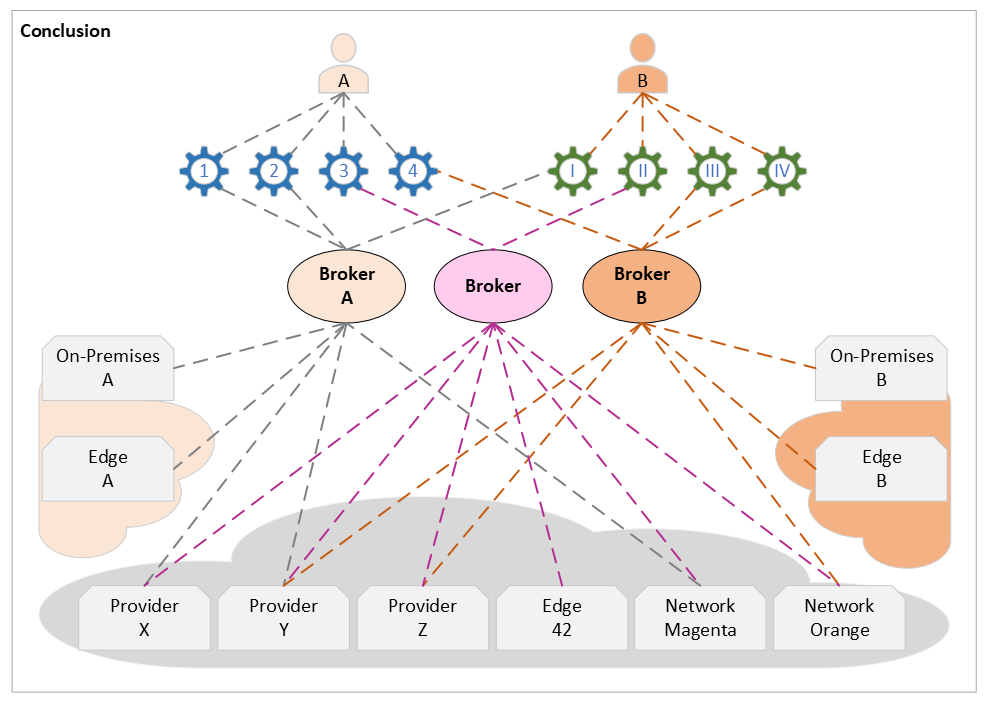}
    \caption{The broker serves as an over-the-top provider for various types of infrastructure.}
    \label{fig:conclusion}
\end{figure}

A resource broker serves as the foundation of effective resource management. By unifying diverse infrastructure resources under a centralized framework, it improves efficiency, enhances scalability, and promotes interoperability. The broker system is an attempt to consistently separate the operation of a service from the provision of the necessary resources. The provision is carried out by defining own offers for required resources and mapping them to the real resources of the underlying providers. These may vary, but the offer remains the same. The proposed approach enables easy scalability, as new infrastructures and basic services can be easily integrated. Working with template chains allows abstraction of real resources as well as their combination into customized offerings to the requesting services.

The diversity of resources from real providers and the possibilities of technologies for developing a new service are unlikely to be covered by the broker. However, during operation, only a well known and highly limited range of configurations exists, which can then be easily distributed across different providers. 

Peaks in demand from the servers can also be transferred to the currently most cost-effective providers, or demands in locations that are not maintained can be serviced. Likewise, access to own infrastructure can be provided to partners by assembling and making suitable offerings for their services accessible through the broker as outlined in Figure \ref{fig:conclusion}.

The broker ultimately serves as a blueprint for such a supporting infrastructure service. At its core, it provides each service with a defined communication channel and data protocol to acquire the necessary resources based on their requirements. The approach of using events and data-driven processes enables easy adaptation to new providers for the required resources.


\section*{Acknowledgements}
Funded by the Bundesministerium für Bildung und Forschung (BMBF, German Federal Ministry of Education and Research) -- 16KISK174K, 16KISK177, 16KISK186.

\balance

\printbibliography

\clearpage

\end{document}